\documentclass[nofootinbib,column,aps,prd,amsmath]{revtex4-2} 
\usepackage{color}
\usepackage{graphicx}
\usepackage{amssymb}
\usepackage{amsmath}
\usepackage{subfig}
\usepackage{hyperref}
\usepackage[utf8]{inputenc}
\usepackage{array}
\usepackage{colortbl}
\usepackage[table]{xcolor}

\begin{document}
	 
\title{Generating ultra-compact neutron stars with bosonic dark matter}

\author{Sarah Louisa Pitz}
\email{pitz@itp.uni-frankfurt.de}
\affiliation{Institut f\"ur Theoretische Physik, Goethe Universit\"at, Max-von-Laue Straße 1, D-60438 Frankfurt am Main, Germany}

\author{J\"urgen Schaffner-Bielich}
\email{schaffner@astro.uni-frankfurt.de}
\affiliation{Institut f\"ur Theoretische Physik, Goethe Universit\"at, Max-von-Laue Straße 1, D-60438 Frankfurt am Main, Germany}

\date{\today}

\begin{abstract}
In this work we investigate the properties of neutron stars admixed with self-interacting scalar bosonic dark matter. The dark matter interaction is described by a generalized $\phi^n$ power-law potential. We perform a stability analysis of these two-fluid objects by studying the onset of the unstable radial modes. We find ultra-compact neutron star-dark matter configurations where the neutron star matter is confined to a core radius of values below $7$ km which is unreachable for pure neutron stars. The total gravitational maximum mass of these ultra-compact configurations can have values of $3.4 \, M_\odot$. With our general ansatz of the power-law potential we show that the compactness of these solutions can be extreme, i.e.\ the compactness is $C = 1/3$ or even larger, making them compact enough to have a light-ring mimicking black holes. 
These ultra-compact objects are stable and possess a dark matter halo while having a hadronic matter core.
With the addition of dark matter to neutron stars recent unusual mass-radius measurements of compact stars can be explained. 
We conclude that apparently contradictory measurements of  neutron star masses and radii could be not only an indication of the presence of dark matter around a the hadronic matter core which is stabilized by the gravitational potential of dark matter but could also serve to disentangle the self-interaction strength of dark matter. Our work points to a stiff equation of state for the dark matter fluid, rather than a soft one.
\end{abstract}

\maketitle

\section{Introduction}\label{intro}
Neutron stars mark the endpoint of stellar evolution for main sequence stars. Next to black holes and white dwarfs they are compact remnants of a supernova explosion. This production mechanisms sets limits on the masses and radii of neutron stars. The minimal mass of a neutron star obtained from a supernova simulation is $M = 1.273 M_\odot$ \cite{PhysRevLett.104.251101} or more recently $M = 1.192 M_\odot$ \cite{Muller:2024aod}. Due to their high densities they offer a unique possibility to study matter under extreme conditions. The properties of these objects strongly depend on the equation of state (EoS), a relation between the energy density and the pressure. Measurements of masses and radii set constraints on the EoS. One requirement is that the maximum mass of a neutron star should exceed of $ 2 \, M_\odot$ 
\cite{NANOGrav:2019jur, Romani_2022}. 
Recently, contradictory measurements of unusual small masses and radii have caught attention, such as HESS J1731-347 with a mass of $M = 0.77 \, M_\odot$ and a radius of $R = 10.4$ km \cite{2022NatAs...6.1444D} as well as XTE J1814-338 with a mass of $M = 1.21 \, M_\odot$ and radius of $R = 7.0$ km  \cite{Kini:2024ggu}. Within purely hadronic models, even when including a phase transition at high densities, it is challenging to describe the properties of these objects, one of them would   
even violate causality. Moreover, the LIGO-Virgo collaboration measured a gravitational wave event where one of the objects is in the mass gap between the lightest black hole and the most massive neutron star, GW230529 \cite{LIGOScientific:2024elc}.  As a remark we would like to point out that results from supernova simulations indicate that the gap can be populated by black holes produced in a fallback supernova \cite{2020ApJ...896...56W}. In this case, dark matter is no longer needed to explain objects that lie in this region. Nevertheless this does not strictly exclude dark matter and it is worth to investigate its impact on a neutron star. Dark matter (DM) can be included in compact stars through different mechanisms such as accretion of dark matter into a neutron star \cite{PhysRevD.82.063531, PhysRevLett.115.111301, PhysRevD.77.023006}, or dark stars that accretes ordinary matter \cite{Meliani_2016, Kamenetskaia:2022lbf}, a merger of a neutron star with a dark star \cite{Diedrichs:2023trk} or the decay of standard model particle into dark matter \cite{PhysRevLett.121.061801, doi:10.1142/S0217751X18440207, Husain_2022}. For dark matter it is possible to form a purely dark star. One possibility are boson stars, self-gravitating spheres, described by complex scalar fields \cite{Colpi:1986ye}. 
They could be build via standard structure formation in the early universe \cite{10.1093/mnras/215.4.575, Chang_2019}. 
Their properties such as masses and radii can resemble those of neutron stars or black holes \cite{10.1093/mnras/staa1878}. Boson stars described by a sufficiently stiff potential, i.e.\ having a stiff equation of state, become ultra-compact with a compactness of $C = GM/R > 1/3$ as shown in our previous work \cite{Pitz:2023ejc}. In this case these configurations, dubbed ultra-compact objects, are compact enough to posses a light-ring, making them potential black hole mimicker. However, in comparison to black holes, boson stars do not have a shadow since light does not interact with dark matter and thus passes though the boson star. By including a significant amount of dark matter into a neutron star one modifies the properties of the star. The simplest approach is to couple the two fluids only gravitationally, i.e.\ without any additional interactions except self-interactions of the separate fluids. In our work we study neutron stars admixed with bosonic dark matter. We use a complex scalar field to model the dark matter component with a stiff self-interaction potential and assume that the dark matter and ordinary matter (OM) fluids only interact via gravitation. There exists a variety of work that has investigate the presence of bosonic dark matter in neutron stars \cite{Rutherford:2022xeb}, especially for the standard $\phi^4$ theory \cite{RafieiKarkevandi:2021hcc, Shakeri:2022dwg, Karkevandi:2024vov, Diedrichs:2023trk, PhysRevD.105.023001, particles7010011} or for vector dark matter \cite{Jockel:2023rrm}. For work investigating fermionic dark matter in neutron stars we refer to \cite{GOLDMAN2013200, PhysRevD.92.123002, Rezaei_2017, PhysRevD.96.083004, Nelson_2019, PhysRevD.99.083008, PhysRevD.102.063028, PhysRevD.105.043013, Ferreira_2023, PhysRevD.108.103016, thakur2023exploringrobustcorrelationsfermionic, 10.1093/mnras/stad3658, BRAMANTE20241, PhysRevD.109.043038, liu2024darkmattereffectsproperties, Vikiaris:2024phe, Koehn:2024gal}
where the second (dark) fluid is modeled as a Fermi gas \cite{Barbat:2024yvi, LI201270, PhysRevD.84.107301}. In a previous work \cite{Zhang:2020dfi} neutron stars with bosonic dark matter were studied with a $\phi^n$ potential up to $n=10$, so that the compactness is less than $C<0.3$. 
Here we extend the study to higher values of the power $n$ \cite{Pitz:2023ejc} so that configurations of pure bosons stars have a compactness of $C > 0.3$. Indeed, we find that such boson stars allow for the possibility of stable ultra-compact neutron stars demonstrating this hitherto unprecedented feature for the case of $n=40$. 
This high power-law potential is stiff enough to exceed a compactness of $C = 1/3$ and to generate ultra-compact solutions. 
Higher orders of the potential do not give a significant increase of the compactness since all solutions go asymptotically to an upper limit, 
which is given by the limit of causality $C = 0.354$ \cite{Pitz:2023ejc}. 

This paper is structured as follows: first we describe the equations of state used for ordinary neutron star matter and dark matter. Then we perform a stability analysis of the combined two-fluid solution of the Tolman-Oppenheimer-Volkoff equation. We presents results for the mass-radius relation, the DM fraction and 
the radii of the neutron star matter and dark matter component delineating the solutions for DM halo and DM core configurations. 
Finally, we discuss our results for masses, radii and compactness in the light of recent measurements of exotic masses and radii of neutron stars. 
We find solutions that fulfill both constraints from HESS J1731-347 and XTE J1814-338 by the virtue of the presence of a stiff DM fluid demonstrating 
that such measurements can serve as a smoking-gun signal for the presence of stiff dark matter EoS. Especially with the upcoming $O5$ run of the Ligo, Virgo and Kagra detectors, the sensitivity could be sufficient to distinguish between light neutron stars, primordial black holes and boson stars \cite{Crescimbeni:2024qrq}.


\section{Theoretical Framework}

\subsection{Tolman-Oppenheimer-Volkoff equations}

The Tolman-Oppenheimer-Volkoff (TOV) equations describe spherical, non-rotating compact objects in hydrostatic equilibrium. By solving the TOV equations one obtains the mass and radius of a compact object. To do so one needs the EoS as an input and fix the central pressure (i.e. the pressure at $r = 0$ where $r$ is the radial component). For the two-fluid case these equations are modified due to the gravitational coupling of the fluids. In this case two EoSs (one for OM and one for DM) are required as well as two central pressures $p_{DM}^c$ and $p_{OM}^c$ which not necessarily have to be the same:
\begin{eqnarray}
    \frac{d p_{OM}}{dr} &=& - \left( p_{OM} + \varepsilon_{OM} \right) \frac{d \nu}{dr}\\
    \frac{d p_{DM}}{dr} &=& - \left( p_{DM} + \varepsilon_{DM} \right) \frac{d \nu}{dr}\\
    \frac{d m_{OM}}{dr} &=& 4 \pi r^2 \varepsilon_{OM} \\
    \frac{d m_{DM}}{dr} &=& 4 \pi r^2 \varepsilon_{DM}
\end{eqnarray}
with 
\begin{equation}
    \frac{d \nu}{dr} = \frac{\left( m_{OM} + m_{DM}\right) + 4 \pi r^3 \left( p_{OM} + p_{DM}\right)}{r (r - 2(m_{OM} + m_{DM}))}
\end{equation}
where $p_{DM}$ is the dark matter pressure, $p_{OM}$ the ordinary matter pressure, $\varepsilon_{OM}$ the EoS for ordinary matter, $\varepsilon_{DM}$ the EoS for dark matter, $r$ the radius, $m_{OM}$ the ordinary matter mass and $m_{DM}$ the dark matter mass. For all our calculation we use natural units, i.e. $\hbar = c = 1$.

\subsection{Dark matter equation of state}

In our work we use the equation of state of our previous work \cite{Pitz:2023ejc}, see also \cite{Zhang:2020dfi}. 
In this model the dark matter is described by complex scalar fields with the Lagrangian
\begin{equation}\label{Lagrangian}
    \mathcal{L} = \partial^\mu \phi^* \partial_\mu \phi + m^2 \phi^* \phi - V
\end{equation}
and a generalized power-law self-interaction potential
\begin{equation}
    V = \frac{\lambda}{2^{n/2}} \left( \phi^* \phi\right)^{n/2}
\end{equation}
with the coupling strength $\lambda$, the bosonic field $\phi$ and the exponent $n$ that can have arbitrary values as long as $n \geq 3$ is fulfilled. The power index $n$ is a direct measurement of the stiffness of the potential. The EoS can be cast in a dimensionless form and is then given by
\begin{equation}
    \varepsilon' = p'^{n/2} + \frac{n + 2}{n - 2}\,p'.
\end{equation}
Here $p'$ and $\varepsilon'$ denote dimensionless quantities with $\varepsilon / \varepsilon' = p/p' = \varepsilon_0$. To obtain dimensionful values one needs to rescale the EoS in the following manner:
\begin{equation}
    \varepsilon = \varepsilon_0 \left( \frac{p}{\varepsilon_0}\right)^{2/n} + \frac{n + 2}{n - 2}\,p
\end{equation}
where $\varepsilon_0 = \lambda \left(\frac{n}{2}-1 \right)m_b^n$ with the boson mass $m_b$. Note that the dimension of $\lambda$ is $[\lambda] = \text{MeV}^{4-n}$ which together with the dimension of the boson mass $[m_b]^n = \text{MeV}^n$ gives an overall dimension of MeV$^4$ for the energy density and the pressure. 

\subsection{Ordinary matter equation of state}

To model the EoS for ordinary matter we use the Baym-Pethick-Sutherland (BPS) equation of state for the outer crust of the neutron star,
for the inner crust the one of Negele-Vautherin \cite{NEGELE1973298}, 
and for the neutron matter equation of state the results from chiral effective field theory $\chi$EFT \cite{Hebeler_2013}. 
To describe the ordinary matter in the core we use the piecewise-polytrope ansatz of \cite{Kurkela:2014vha}:
\begin{eqnarray}
    \varepsilon = \frac{1}{\Gamma - 1}p + \left( \mu_0\, n_0 - \frac{\Gamma}{\Gamma - 1} p_0\right) \left(\frac{p}{p_0}\right)^{1/\Gamma}
\end{eqnarray}
where $\Gamma$ is a parameter with $\Gamma > 1$, $\mu_0$ is the matching chemical potential, $n_0 = n(\mu_0)$ is the number density and $p_0 = p(\mu_0)$. All the parameters used can be found in table \ref{ValuesFirstMono} and \ref{ValuesSecondMono}. We choose the parameter sets EoS1 and EoS2 of
\cite{Kurkela:2014vha} as limiting cases of soft ans stiff equations of state, both of them fulfill the $2 M_\odot$ mass limit \cite{NANOGrav:2019jur, Romani_2022}. 
Furthermore, EoS1 is sufficiently soft to satisfy the tidal deformability constraints from GW170817 \cite{LIGOScientific:2017vwq}.

\begin{table}
\centering
\begin{tabular}{|p{2cm}|p{2cm}|p{2cm}|p{2cm}|} 
 \hline
   &$M_{max} [M_\odot]$ & $R_{max}$ [km] & $C_{max}$  \\ 
  \hline
EoS1 & $2.03$ & $10.4$ & $0.29$  \\ 
 \hline
  EoS2 & $2.45$ & $13.3$ & $0.27$ \\ 
 \hline
\end{tabular}
 \caption{Values for the maximum mass, the corresponding radius and the maximum compactness of EoS1 and EoS2.}
 \label{ValuesEoS1and2}
\end{table} 

\begin{table}
\centering
\begin{tabular}{|p{2cm}|p{2cm}|p{2cm}|p{2cm}|p{2cm}|p{2cm}|} 
 \hline
  &$\Gamma$ & $\mu_0$ [MeV] & $n_0 / n_{sat}$ & $p_0$ [MeV/fm$^3$] & $K$ [$\text{MeVfm}^{3(\Gamma-1)}$] \\ 
  \hline
EoS1 & $3.192$ & $0.9657$ & $1.1$ & $2.136$ & $553.6$ \\ 
 \hline
  EoS2 & $4.021$ & $0.9775$ & $1.1$ & $3.542$ & $3828.6$\\ 
 \hline
\end{tabular}
 \caption{Parameters for the first monotropes for EoS1 and EoS2. Here $n_0$ is given in units of the saturation density $n_{sat} = 0.16$ fm$^{-3}$.}
 \label{ValuesFirstMono}
\end{table} 

\begin{table}
\centering
\begin{tabular}{ |p{2cm}|p{2cm}|p{2cm}|p{2cm}|p{2cm}|p{2cm}|} 
 \hline
  &$\Gamma$ & $\mu_0$ [MeV] & $n_0 / n_{sat}$ & $p_0$ [MeV/fm$^3$] & $K$ [$\text{MeVfm}^{3(\Gamma-1)}$]\\ 
  \hline
  EoS1 & $1.024$ & $1.653$ & $5.9$ & $455.3$ & 482.9\\ 
 \hline
  EoS2 & $1.195$ & $1.351$ & $2.7$ & $129.7$ & 343.6\\ 
 \hline
\end{tabular}
 \caption{Parameters for the second monotropes for EoS1 and EoS2. Here $n_0$ is given in units of the saturation density $n_{sat} = 0.16$ fm$^{-3}$.}
 \label{ValuesSecondMono}
\end{table}

\subsection{Stability analysis}

We perform a stability analysis for the two-fluid objects in the manner of Hippert et al. \cite{Hippert:2022snq}, which was also used to study neutron stars with fermionic dark matter \cite{Barbat:2024yvi}. For DM masses in the range of $1$ GeV the results of their analysis and ours are almost identical, however they also considered DM masses above $1$ GeV and below $200$ MeV which give different results for the stability regions and the masses and radii of the objects. For $m_b > 1$ GeV the objects are of the size of intermediate-mass black holes and for  $m_b< 200$ MeV in the mass range of planets. These masses were not of interest for our calculations since we want to study objects with the characteristic size of a neutron star. 
For the stability analysis one considers radial perturbations of the density. The onset of the unstable modes for the two-fluid case is characterized by 
\begin{eqnarray}\label{HippertStabCon}
    \begin{pmatrix}
        \delta N_{OM} \\
        \delta N_{DM}
    \end{pmatrix} = 
    \begin{pmatrix}
        \partial N_{OM}/\partial\varepsilon_{OM}^c & \partial N_{OM}/\partial\varepsilon_{DM}^c\\
        \partial N_{DM}/\partial\varepsilon_{OM}^c & \partial N_{DM}/\partial\varepsilon_{DM}^c
    \end{pmatrix}
    \begin{pmatrix}
        \delta \varepsilon_{OM}^c \\
        \delta \varepsilon_{DM}^c
    \end{pmatrix} = 0
\end{eqnarray}
where $N_{OM}$ is the total OM and $N_{DM}$ the DM number, $\varepsilon^c$ denotes the central energy density of the OM or DM. Making use of the stationarity of the particle numbers given by equation (\ref{HippertStabCon}) one can diagonalize the matrix and obtain two eigenvalues $\kappa_A$ and $\kappa_B$. Stable configurations are characterized by both eigenvalues being positive: $\kappa_A > 0$ and $\kappa_B > 0$.
For this purpose we need to calculate the total baryon and total DM number $N_{OM}$ and $N_{DM}$, respectively. For the baryon number we have used the BPS, Negele-Vautherin and $\chi$EFT EoS. For the polytropes we integrated the number densities $n$
\begin{eqnarray}\label{totNumbs}
    \frac{dN}{dr} = 4 \pi \left( 1 - \frac{2 m}{r}\right)^{-1/2} n \,r^2 dr.
\end{eqnarray}
where
\begin{eqnarray}
    n_{polytrope} = \left( \frac{p_{OM}}{K}\right)^{1/\Gamma}.
\end{eqnarray}
Here $K$ is a constant that has different values for the two monotropes which can be found in table \ref{ValuesFirstMono} and \ref{ValuesSecondMono}. The dark matter number density is obtained by the zeroth component of the Noether current:
\begin{eqnarray}
    j^\mu = \frac{\partial \mathcal{L}}{\partial(\partial_\mu \phi^*)} i \phi^* - \frac{\partial \mathcal{L}}{\partial(\partial_\mu \phi)} i \phi
\end{eqnarray}
with the Lagrangian from equation (\ref{Lagrangian}) and $\phi = \phi_0\, e^{-i\omega t}$. We find:
\begin{eqnarray}
    j^0 = 2 \left( \frac{2^{n/2}}{\lambda} \left(\frac{n}{2}-1 \right) p_{DM}\right)^{2/n} \times \left[ m_b^2 + \frac{\lambda}{2^{n/2}} \frac{n}{2} \left( \frac{2^{n/2}}{\lambda} \left(\frac{n}{2} -1 \right) p_{DM}\right)^{1 - \frac{2}{n}}\right]^{1/2}.
\end{eqnarray}
In the manner of equation (\ref{totNumbs}) one obtains the total dark matter number $N_{DM}$ by setting $n=j^0$. A more extensive analysis can be found in \cite{Caballero:2024qtv}.


\section{Results}

For our calculations we have used two different EoSs to model the ordinary matter and two different EoSs to model the dark matter. The ones for the ordinary matter are EoS1 and EoS2, both piecewise polytropes as presented above. In the case of the dark matter we are using two different values of the exponent $n$, namely $n=4$ and $n=40$. We chose $n=4$ since it is used as a standard self-interaction potential for boson stars and thus enables to compare our results to previous ones such as \cite{RafieiKarkevandi:2021hcc,Shakeri:2022dwg,Karkevandi:2024vov, Diedrichs:2023trk,PhysRevD.105.023001,Shakeri:2022dwg,Karkevandi:2024vov,particles7010011}. 
We are furthermore studying the case of $n$ being much larger than $n=4$ because we want to explore the possibility of ultra-compact neutron stars, e.g. objects with a compactness of $C \geq 1/3$. The potential with $n=40$ is a stiff potential with a corresponding stiff equation of state that results in ultra-compact boson stars, so it is of high interest to explore its impact on a two-fluid system such as neutron stars with admixed dark matter. 
We want to study if the gravitational interaction between the two fluids is already sufficient to generate neutron stars with extraordinary small masses and radii. As demonstrated in a previous work higher orders of the potential do not give a significant increase of the compactness for boson stars \cite{Pitz:2023ejc}. 
The upper limit is here given by the limit of causality $C = 0.354$ where the speed of sound is equal to the speed of light $c_s^2 = 1$. Already with $n=40$ we exceed $C=1/3$ and have a compactness of $C=0.34$. For $\phi^4$ the maximum compactness is $C=0.16$ \cite{PauAmaro-Seoane_2010} which is per definition not high enough to generate ultra-compact objects.  

\subsection{Results for the mass-radius curves}

In the following we present masses and radii where each mass-radius curve is calculated for a fixed value of the dark matter central pressure $p_{DM}^c$. 
Low DM pressures are indicated by the darker colors and higher ones by the lighter colors. The solid lines in the plots mark the stable configurations and the dashed ones the unstable ones in the mass-radius curve.

\begin{figure*}
   \centering
    \resizebox{1.0\textwidth}{!}{
    \begin{tabular}{cc}
        \subfloat[]{%
            \includegraphics[width=0.5\textwidth]{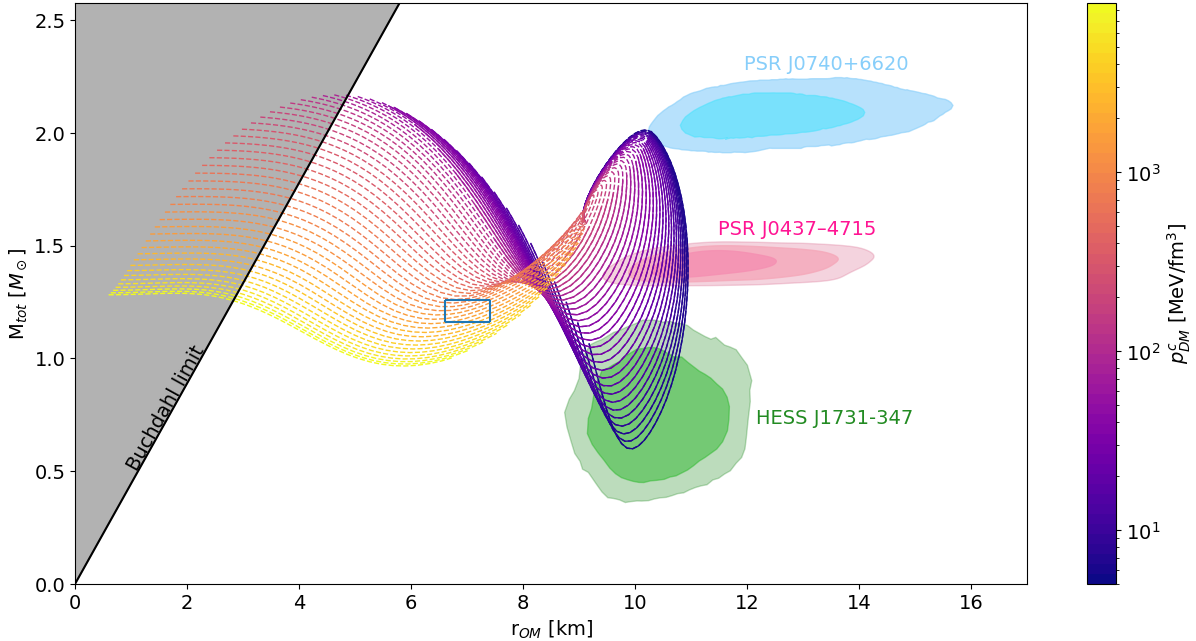}
            \label{fig:MReos1a}
        } &
        \subfloat[]{%
            \includegraphics[width=0.5\textwidth]{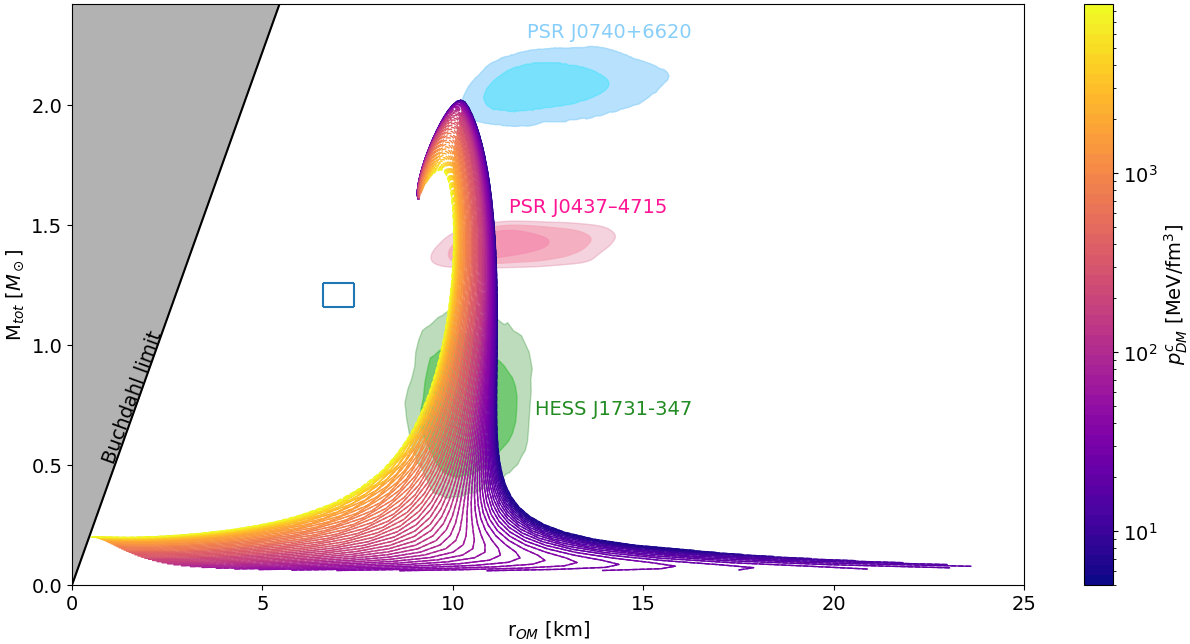}
            \label{fig:MReos1b}
        } \\
        \subfloat[]{%
            \includegraphics[width=0.5\textwidth]{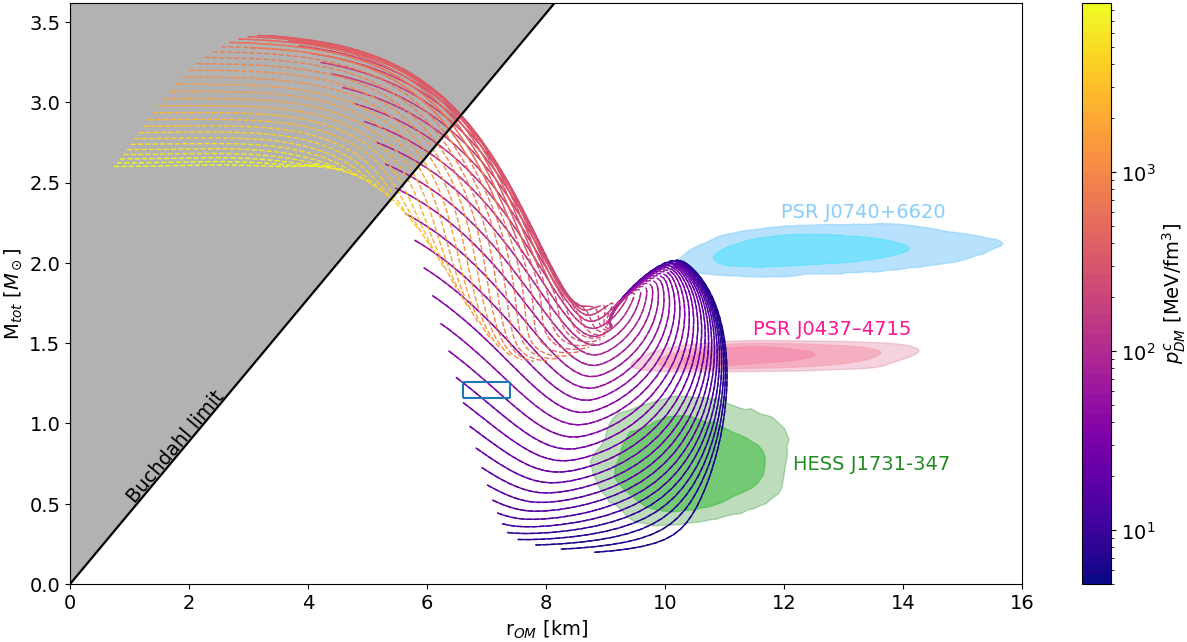}
            \label{fig:MReos1c}
        } &
        \subfloat[]{%
            \includegraphics[width=0.5\textwidth]{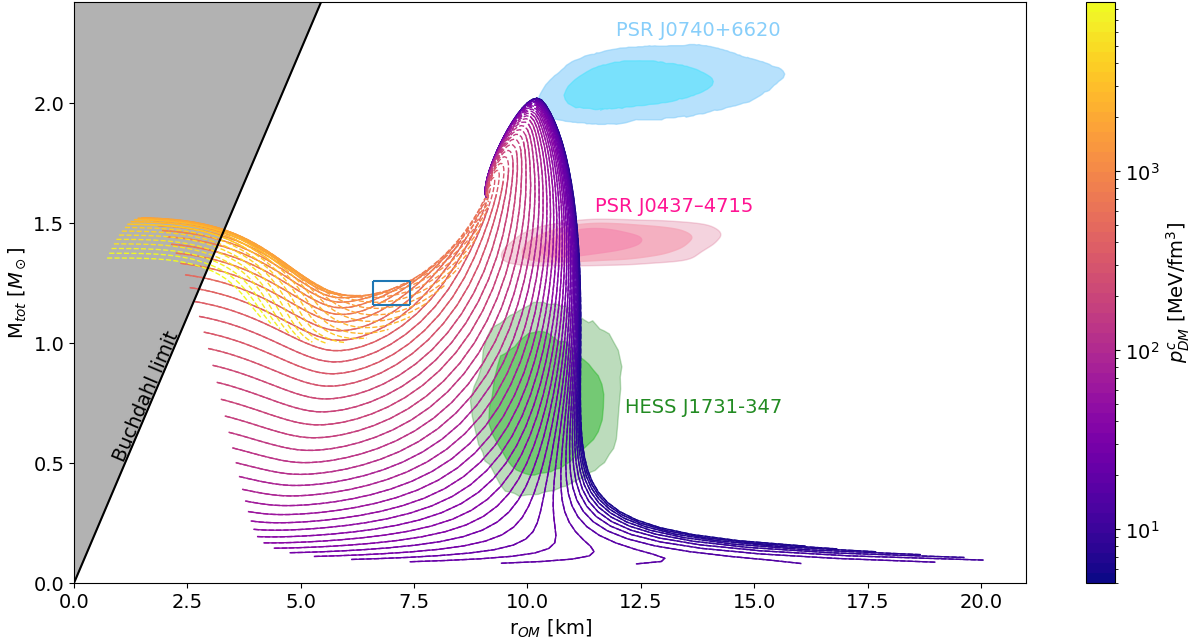}
            \label{fig:MReos1d}
        }
    \end{tabular}}
    \caption{Mass-radius curves for EoS1 together with DM with a scalar potential of $n=4$ (upper panel) and of $n=40$ (lower panel). 
    (a) the case for a boson mass of $m_b = 300$ MeV, (b) for $m_b = 1000$ MeV, (c) for $m_b = 200$ MeV and (d) again for $m_b = 300$ MeV. 
    Color-coded is the DM central pressure in MeV/fm$^3$. The solid lines represent the stable configurations and the dashed ones the unstable part. Included are the NICER constraints from PSR J0740+6620 \cite{2024arXiv240614467D, 2024arXiv240614466S}
 (blue curves), PSR J0437–4715 \cite{2024ApJ...971L..20C} (pink curves),  HESS J1731-347 \cite{2022NatAs...6.1444D} (green curves), and  XTE J1814-338 \cite{Kini:2024ggu}(blue rectangle). Note that the comparison with the data involves the radius of ordinary matter as the radius measurements involve photons not gravity alone.}
    \label{fig:MReos1}
\end{figure*}

Our free parameters are the boson mass $m_b$ and the stiffness of the potential, indicated by $n$. For each $n$ we choose a mass that gives results that fulfill most of the constrains, especially the ones for HESS J1731-347 and XTE J1814-338. For $n=4$ this turns out to be $m_b = 1000$ MeV and $m_b = 300$ MeV and for $n=40$ we chose $m_b = 200$ MeV and $m_b = 300$ MeV. All our curves with EoS1 satisfy the HESS measurement as well as the updated constraints on PSR J0437–4715. Furthermore, we find that our results for a stiff potential are also able to describe the XTE J1814-338 data with $M = 1.21 \pm 0.05 \, M_\odot$ and $R = 7.0 \pm 0.4$ km, which could be an indication for the presence of dark matter in this compact object, since a neutron star consisting only of ordinary matter would violate causality.

\begin{figure*}
    \centering
    \resizebox{1.0\textwidth}{!}{
    \begin{tabular}{cc}
        \subfloat[]{%
            \includegraphics[width=0.5\textwidth]{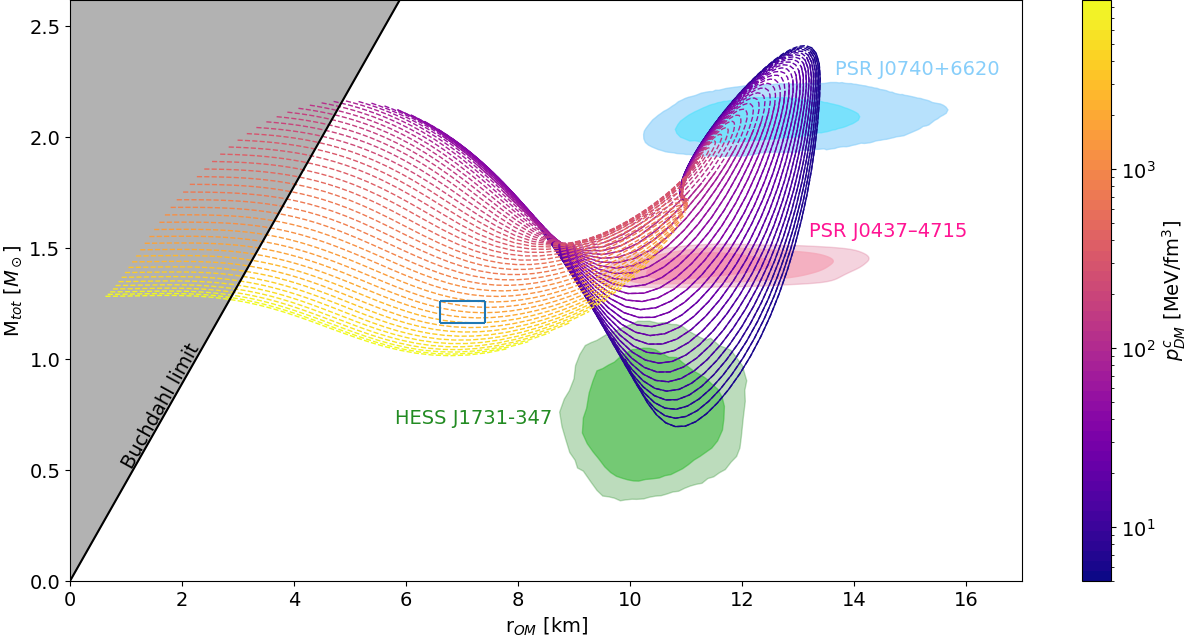}
            \label{fig:MReos2a}
        } &
        \subfloat[]{%
            \includegraphics[width=0.5\textwidth]{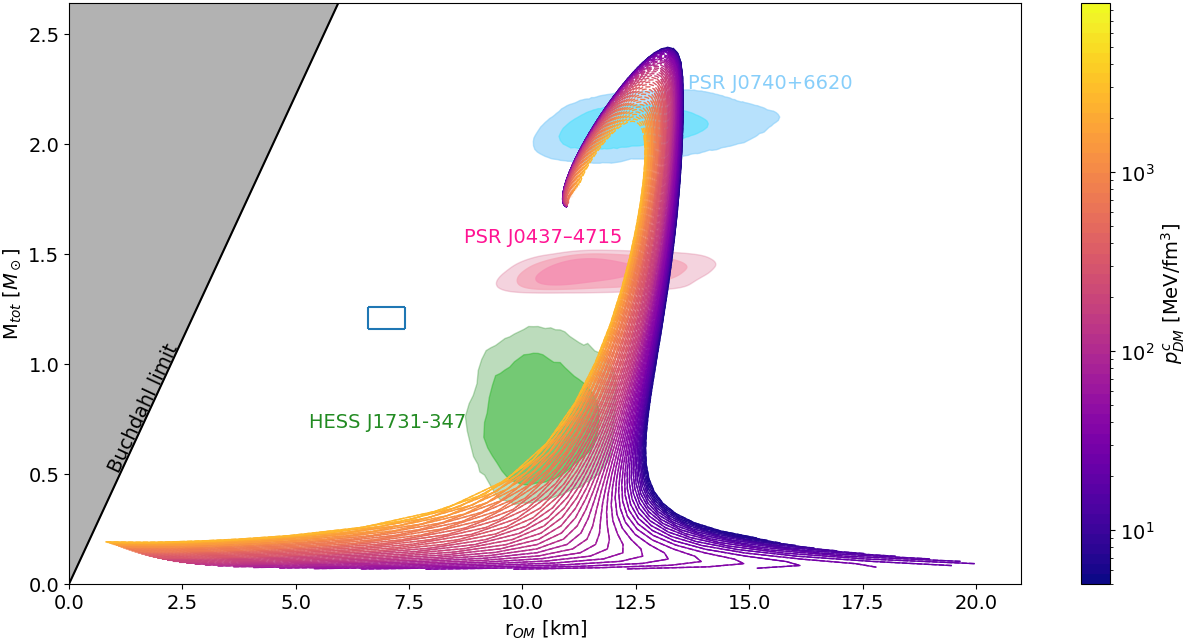}
            \label{fig:MReos2b}
        } \\
        \subfloat[]{%
            \includegraphics[width=0.5\textwidth]{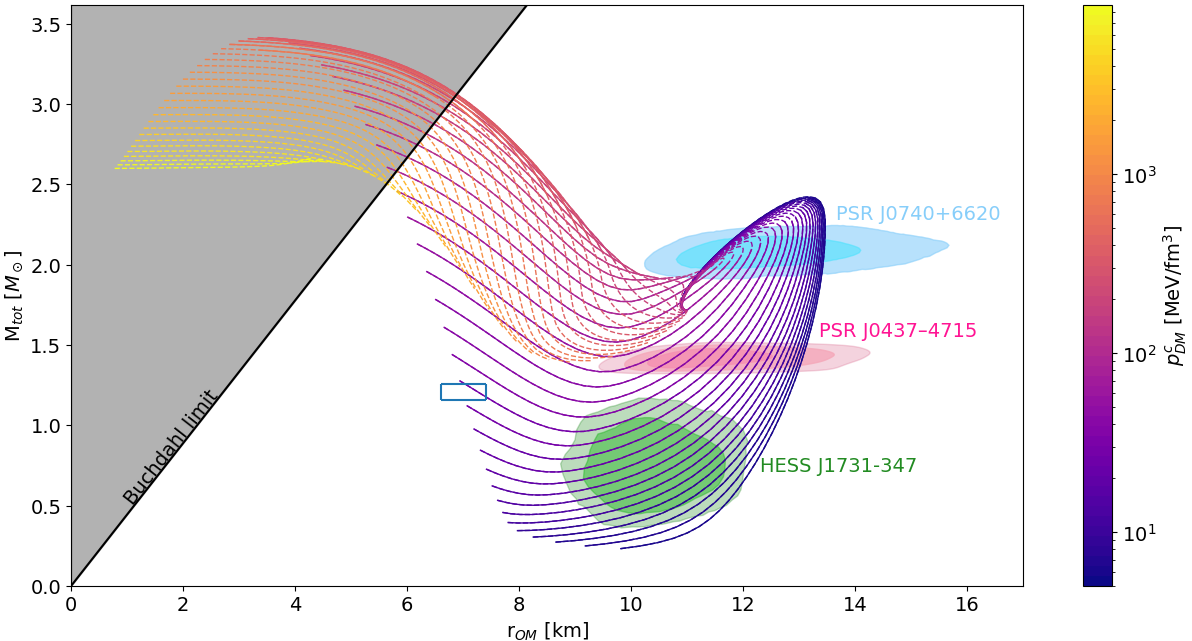}
            \label{fig:MReos2c}
        } &
        \subfloat[]{%
            \includegraphics[width=0.5\textwidth]{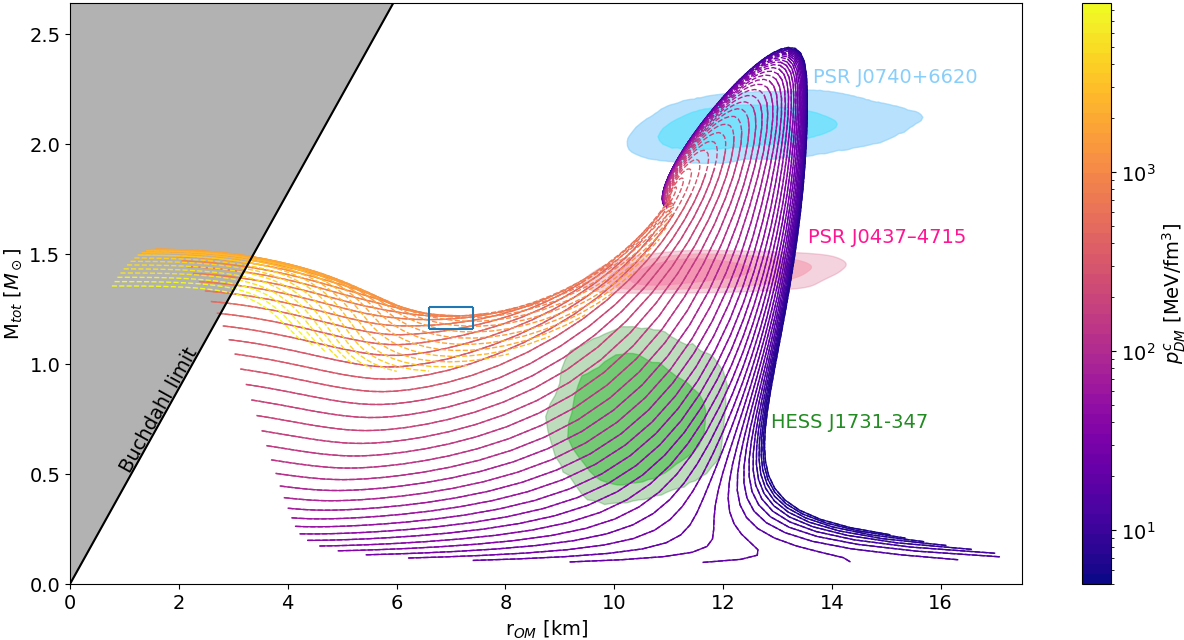}
            \label{fig:MReos2d}
        }
    \end{tabular}}
    \caption{Mass-radius curves for EoS2 for $n=4$ (upper panel) and $n=40$ (lower panel). The labeling is the same as in figure \ref{fig:MReos1}.}
    \label{fig:MReos2}
\end{figure*}

In figure \ref{fig:MReos1} we show the mass-radius curves with the total mass $M_{tot} = M_{OM}+M_{DM}$ and the OM radius. In our work we carefully distinguish between the gravitational radius and the photometric radius, which is the radius that is seen by photons. We would like to stress that the Buchdahl limit $C = 4/9$, indicating the maximum compactness of a static, spherically symmetric star where the energy density does not increase outwards, is only apparently violated since we show the photometric and not the gravitational radius. The same applies for figure \ref{fig:MReos2}. In figure \ref{fig:MReos1a} we show the mass-radius curves of a $\phi^4$ potential with a boson mass of $m_b = 300$ MeV. We find that with increasing DM central pressure the curves become more and more unstable. However we are able to describe the HESS data with a DM central pressure in the range of $1000-2000$ MeV/fm$^3$. Figure \ref{fig:MReos1b} shows the same potential but with a boson mass of $m_b=1000$ MeV. For this mass the influence of the DM is lower compared to figure \ref{fig:MReos1a} but again the curves with a high DM central pressure tend to result more likely in unstable solutions. With this combination of potential and DM mass we also describe the HESS data but not XTE J1814-338. This changes for $n=40$, here we find stable solutions for both XTE J1814-338 and HESS J1731-347. Moreover, for $m_b=200$ MeV, see figure \ref{fig:MReos1c}, we find an increase of the maximum stable masses. Here we have configurations with a mass above $3 M_\odot$ and a DM central pressure in the range of $5000$-$6000$ MeV/fm$^3$. These objects could serve as an explanation for the recent GW230529 measurement where one object was in the mass gap between $2.5$-$4.5 M_\odot$ \cite{LIGOScientific:2024elc}. Moreover we want to stress that these objects have high total masses, while having small OM radii. In order to find these objects we note that a $\phi^4$ potential is not sufficiently stiff since all the high-mass configurations are unstable. For the case of $m_b=300$ MeV depicted in figure~\ref{fig:MReos1d} the mass-radius curves lie within the constraints of both, XTE J1814-338 and HESS J1731-347. For all the curves shown in figure~\ref{fig:MReos1} we find in each scenario stable neutron stars with a mass below $1.4 M_\odot$ and radius below $8$ km. With purely hadronic EoSs it is not possible to populate this region without violating causality. We point out that XTE J1814-338 lies exactly in this area and thus could be interpreted as an indication for the presence of DM. Furthermore is shows that these small masses and radii, especially the ones that result in ultra-compact configurations, require a rather stiff than soft DM EoS. 

In figure \ref{fig:MReos2} we display the mass-radius curves for EoS2. As in figure \ref{fig:MReos1} we show the total mass and the OM radius. In figure \ref{fig:MReos2a} we show the case for $\phi^4$ and a DM mass of $m_b=300$ MeV. Similar to figure \ref{fig:MReos1a} we find no stable solutions for XTE J1814-338, however all the other constrains, namely  PSR J0437–4715, HESS J1731-347 and PSR-J0740+6620, are well described by our model. The same is valid for the case of $m_b=1000$ MeV, see figure \ref{fig:MReos2b}. In this case the objects that lie within the HESS data are slightly shifted to a higher DM central pressure. Figure \ref{fig:MReos2c} shows the mass-radius curves for $n=40$ and a mass of $m_b=200$ MeV. Also for this EoS we find that this scenario generates extremely high masses of $M > 3 M_\odot$ and fulfills all constrains, where the configurations within the HESS J1731-347 and XTE J1814-338 measurements lie in the low $p_{DM}^c$ region. This changes slightly for figure \ref{fig:MReos2d}, where one can describe the HESS data with a low central pressure, but not the data of XTE J1814-338. For this DM mass we need a DM central pressure above $6000$ MeV/fm$^3$ to get in the latter mass-radius region. 
For both EoS we only find stable solutions for XTE J1814-338 in the case of $n=40$, which could imply that the self-interaction of DM is rather stiff than soft.

\subsection{Results for the compactness and radii}

We investigate below the pressure parameter-space in order to make statements about the compactness $C = M/R$ (with $G=1$), the amount of DM, and check for configurations with a DM halo or a DM core. To determine the amount of DM present we used the ratio $M_{DM} / M_{tot}$, where $M_{tot} = M_{DM} + M_{OM}$ is the total mass, while in order to distinguish between DM halo and core we used the ratio of the radii $r_{DM} / r_{OM}$. Both radii are determined by the condition $p_i = 0$ where $i$ represent either DM or OM. We consider the following three cases: $r_{DM} / r_{OM} < 1$ which is the case for a DM core, $r_{DM} / r_{OM} = 1$ where both radii are the same, and the case is $r_{DM} / r_{OM} > 1$ where the DM forms a halo around the object. 

\begin{figure}
    \resizebox{1.0\textwidth}{!}{
    \begin{tabular}{cc}
        \subfloat[]{%
            \includegraphics[width=0.5\textwidth]{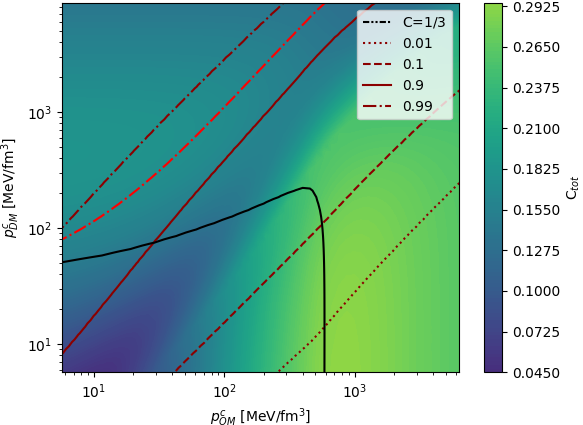}
            \label{fig:ContourEoS1a}
        } &
        \subfloat[]{%
            \includegraphics[width=0.5\textwidth]{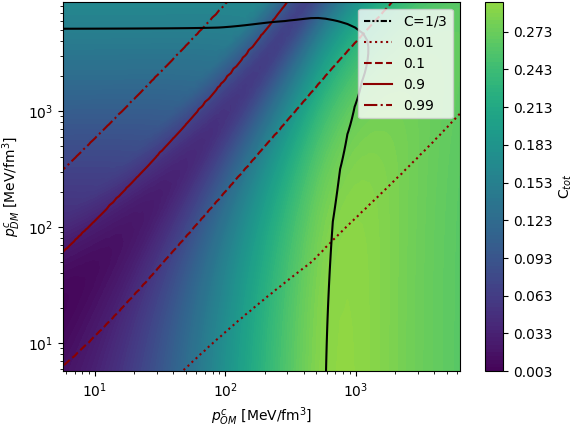}
            \label{fig:ContourEoS1b}
        } \\
        \subfloat[]{%
            \includegraphics[width=0.5\textwidth]{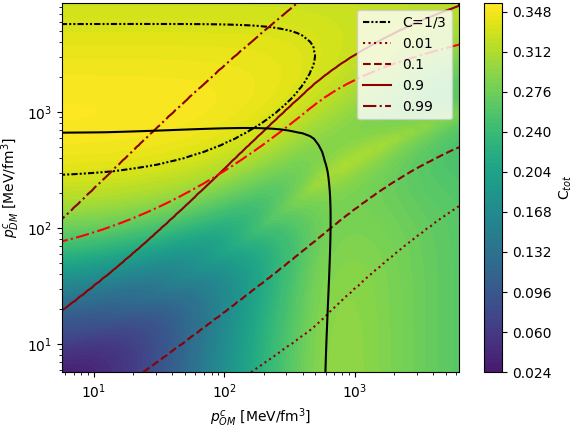}
            \label{fig:ContourEoS1c}
        } &
        \subfloat[]{%
            \includegraphics[width=0.5\textwidth]{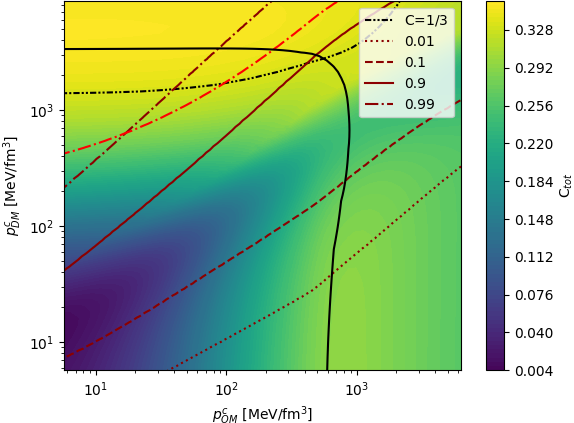}
            \label{fig:ContourEoS1d}
        }
    \end{tabular}}
    \caption{Contour plots of EoS1 together with $n=4$ (upper panel) and $n=40$ (lower panel). The y-axis corresponds to the DM central pressure in MeV/fm$^3$, the x-axis to the OM central pressure, also in MeV/fm$^3$ and color-coded is the total-compactness $C_{tot} = M_{tot} / r$ where $r$ is the gravitational radius of the star. Different amounts of DM are shown as indicated by the legend. (a) corresponds to a boson mass of $m_b = 300$, (b) to $m_b = 1000$ MeV, (c) to $m_b = 200$ MeV and (d) to $m_b = 300$ MeV. The solid black line is the stability line, e.g. all configurations underneath the line are stable, above unstable. The red dashed-dotted line indicates the Buchdahl limit for the pseudo-compactness $C(r_{OM}) = M_{tot} / r_{OM} = 4/9$.}
    \label{fig:ContourEoS1}
\end{figure}

In figure \ref{fig:ContourEoS1} we show contour-plots for the DM and OM central pressures with the compactness. We show four different cases of the DM fraction ($1 \%, 10 \%, 90 \%$ and $99 \%$) as well as the line for a constant compactness of $C=1/3$. Note that the latter line is not visible in every plot, since it is not reached for the case of $n=4$. 
The result of our stability analysis is depicted by the black solid line. One has to read the plots the following way: all the configurations underneath the black line are stable and the ones above are unstable. The Buchdahl limit for the pseudo-compactness $C(r_{OM}) = M_{tot} / r_{OM} = 4/9$ is shown by the red dashed-dotted line. All configurations above this line seem to violate this upper bound. However, this is only valid for the radius of the OM fluid, not for the whole star. In figure \ref{fig:ContourEoS1a} we show the results for $n=4$ and $m_b=300$ MeV. Here the stable region is small compared to figure \ref{fig:ContourEoS1b} and we do not find stable solutions with a DM fraction of $99 \%$. The maximum compactness is $0.29$ which is due to the higher compactness allowed by EoS1, since the $\phi^4$-potential alone gives only a compactness of at most $C=0.16$ \cite{PauAmaro-Seoane_2010}. The same maximum compactness is valid for figure \ref{fig:ContourEoS1b} for the case $m_b=1000$ MeV. For this combination we have a bigger stability window, with stable solutions that have a DM fraction of $99 \%$. In figure \ref{fig:ContourEoS1c} we find configurations where the compactness is $C=1/3$ and even higher. The area of interest is between the dashed-dot-dotted and the stability line, since these are the ultra-compact neutron stars that are stable. These compact stars require a high DM fraction of more than $90 \%$. In these cases one has a boson star with a neutron star in the core. In comparison, for $m_b=300$ MeV, see figure \ref{fig:ContourEoS1d}, we find ultra-compact stable solutions with a lower DM fraction, since $M_{DM}/M_{tot} = 0.9$ is found to be within the $C\geq 1/3$ region. In both cases the ultra-compact neutron stars have a low OM central pressure of only a few MeV/fm$^3$ up to approximately $200$ MeV/fm$^3$.  These pressures are so low that is highly unlikely that there will be a phase transition to quark matter. 

\begin{figure*}
    \centering
    \resizebox{1.0\textwidth}{!}{
    \begin{tabular}{cc}
        \subfloat[]{%
            \includegraphics[width=0.5\textwidth]{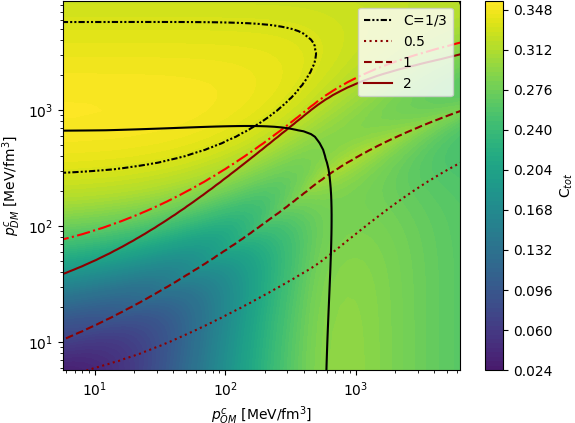}
            \label{fig:RadiiRatioa}
        } &
        \subfloat[]{%
            \includegraphics[width=0.5\textwidth]{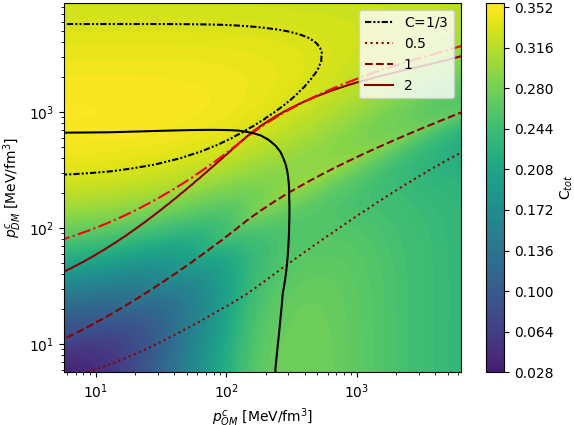}
            \label{fig:RadiiRatiob}
        }
    \end{tabular}}
    \caption{Contour-plots with the DM central pressure on the y-axis and the OM central pressure on the x-axis. Color-coded is the compactness, where the dashed-dotdotted black line corresponds to a constant compactness of $C = 1/3$ and the dashed-dotted red line to a pseudo-compactness $C(r_{OM}) = M_{tot} / r_{OM} = 4/9$. The brown lines represent the ratio of the DM and OM radius: $r_{DM} / r_{OM}$. Here the dotted line is a ratio of $0.5$, which means the DM radius is only half the size of the OM radius, e.g. these configurations posses a DM core, the dashed line corresponds to a ratio of $1$ where both radii are equal and finally the solid line indicates that the DM radius is twice the size of the OM one. In this figure we only considered $\phi^{40}$ since $n=4$ does not give ultra-compact solutions. Both plots are calculated with a boson mass of $m_b = 200$ MeV, the left one is for EoS1 and the right one for EoS2.}
    \label{fig:RadiiRatio}
\end{figure*}

In figure \ref{fig:RadiiRatio} we show the maximum compactness and the regions where the star possesses a DM core or halo, respectively. The border is marked by the dashed brown line that corresponds to a ratio $r_{DM}/r_{OM} = 1$. All the configurations above this line have a DM halo and below a DM core. In all cases studied we find that the ultra-compact objects are in the region above the border line, in fact, they have a DM halo with a DM radius that is at least twice as big as the OM radius. Such a compact star would be visible as one with an unusual small neutron star with a radius of about $2$ - $8$ km such as XTE J1814-338. In figures \ref{fig:MReos1c} and \ref{fig:MReos1d} the configurations with a DM halo correspond to the part of the curves on the left-hand side of the minimum. This can be seen if one plots the total radius instead of the OM one, that the mass-radius curves bend back, meaning that the mass and the radius increase again. Figures \ref{fig:RadiiRatioa} and \ref{fig:RadiiRatiob} are quite similar, nevertheless we find that for EoS2 a DM radius that is approximately twice the size of the OM one is sufficient to generate ultra-compact objects, whereas for EoS1 it needs to be more than two times larger. However for both EoSs the stable stars with $C>1/3$, so the ones that lie between the line for $C=1/3$ and the stability line, all have a DM radius that is more than twice the size of the OM one. Together with results from figures \ref{fig:ContourEoS1} 
we see that these ultra-compact neutron stars all have a DM halo with a total DM fraction above $90 \%$.
All the values for the DM and OM central pressures as well as the range of DM fraction needed to describe HESS J1731-347 and XTE J1814-336, can be found in table \ref{tab:summary}. Note that we only show the values for the curves shown in figures \ref{fig:MReos1} and \ref{fig:MReos2}.
The constraints from HESS and from NICER can be described in all cases. For our chosen EOSs we need only a small fraction of DM to be compatible with the HESS data.
We point out that the HESS data can be also described without any DM with the soft EoS1 which is not shown in the figures. 
The data for XTE is more constraining and a large fraction of DM is needed
to be compatible with the mass-radius constraint. For the case $m_b=300$~MeV also a high pressure of DM is needed while for the case $m_b=200$~MeV a smaller pressure is sufficient.

\begin{table}
    \centering
    \begin{tabular}{|c|c|c|c|c|c|c|c|c|c|c|}
    \hline
         OM EoS & $n$ & $m_b$ [MeV] & $p_{DM}$ [MeV/fm$^3$] @XTE & $p_{DM}$ [MeV/fm$^3$] @HESS & $f_{DM}$ @XTE & $f_{DM}$ @HESS & XTE & HESS & J0437 & J0740\\
         \hline
         $1$ & $4$ & $300$ & - & $5$ - $30$ & - & $(4.5$ - $86)\, \%$ & unstable & $\checkmark$ & $\checkmark$ & $\checkmark$\\
         \hline
         $1$ & $4$ & $1000$ & - & $5$ - $8700$ &  - & $(2.6$ - $34)\, \%$ & X & $\checkmark$ & $\checkmark$ & $\checkmark$\\
         \hline
         $1$ & $40$ & $200$ & $ 30$ - $35$ & $5$ - $35$ & $(84$ - $90)\, \%$ & $(17$ - $35)\, \%$ & $\checkmark$ & $\checkmark$ & $\checkmark$ & $\checkmark$\\
         \hline
         $1$ & $40$ & $300$ & $590$ - $980$ & $5$ - $160$ & $(51$ - $57)\, \%$ & $(1.9$ - $25)\, \%$ & $\checkmark$ & $\checkmark$ & $\checkmark$ & $\checkmark$\\
         \hline 
         \hline 
         $2$ & $4$ & $300$ & - & $5$ - $25$ & - & $(44 - 53)\, \%$ & unstable & $\checkmark$ & $\checkmark$ & $\checkmark$ \\
         \hline
          $2$ & $4$ & $1000$ & - & $ 100$ - $3500$ & - & $(14 - 42)\, \%$ & X & $\checkmark$ & $\checkmark$ & $\checkmark$\\
         \hline
         $2$ & $40$ & $200$ & $26$ & $5$ - $30$ & $89 \, \%$ & $(17$ - $63)\, \%$ & $\checkmark$ & $\checkmark$ & $\checkmark$ & $\checkmark$\\
         \hline
         $2$ & $40$ & $300$ & $450$ - $590$ & $23$ - $160$ & $ 60\, \%$ & $(15$ - $48)\, \%$ & $\checkmark$ & $\checkmark$ & $\checkmark$ & $\checkmark$\\
         \hline
    \end{tabular}
    \caption{Values for the dark matter and ordinary matter central pressures and the dark matter fractions $f_{DM}$ needed to reach the masses and radii reported for HESS J1731-347 and XTE J1814-338 for all the equations of states used. The equations of state satisfying the mass-radius constraints from the NICER measurements are also indicated by ticks.}
    \label{tab:summary}
\end{table}

\section{Summary}

We have studied the properties of neutron stars with bosonic dark matter allowing for ultra-compact configurations. 
For the dark matter interaction we choose a complex scalar field with a stiff self-interaction potential $\phi^n$ where one can determine the stiffness by varying the exponent (the higher $n$ the stiffer the potential, i.e.\ the stronger the energy increases with the field strength). For our studies we choose $n=4$ for comparison and $n=40$. We selected those values because $\phi^4$ is a commonly used self-interaction potential and thus gives the possibility to compare our results to previous work. The case $n=40$ is of interest since it is sufficiently stiff to generate ultra-compact boson stars. For the OM we used two piece-wise polytropic EoSs, namely EoS1 and EoS2 which delineate the range of soft and stiff ordinary matter equations of state being compatible with a maximum mass of $2\,M_\odot$. 
We performed a stability analysis of our results by looking at radial density oscillations involving both fluids. The unusual mass and radius measurements of HESS J1731-347 and XTE J1814-338 could be described  with the chosen approach where we are able to find stable solutions in the mass-radius region of these measurements suggesting the presence of DM. We investigated the possible parameter space of the DM and OM central pressures. We find that the configurations with a compactness $C > 1/3$, i.e. the ultra-compact neutron stars, have a DM fraction of about $\gtrsim 90 \%$ and high DM central pressures. These high compactnesses are challenging to reach with a hadronic EoS only. For each case of dark matter fraction studied ($0.1 \%$, $10 \%$, $90 \%$ and $99 \%$) we find stable solutions, especially also solutions with a compactness higher or equal $C = 1/3$. These are of particular interest since these objects are so compact that they have a light-ring at or even above their surface, making them black hole mimicker. Apart from this we found that these ultra-compact neutron stars have in any case a DM halo, not a DM core. Here the DM radius is about twice the size of the OM radius. In this case we would observe a unusually small neutron star with a radius of only $2$ - $8$ km and a mass below $1.5 \,M_\odot$ with a light-ring placed above its surface. 

We conclude that the introduction of a scalar bosonic field with a stiff self-interaction potential results in exceptional properties of neutron stars with dark matter. The presence of DM has drastic implications on the OM radius and mass of the star and thus also on the compactness. We found that with this two-fluid description we are able to generate ultra-compact and unusually light and small neutron stars which could be an explanation for exotic objects such as HESS J1731-347 and XTE J1814-338. These measurements could therefore be a smoking-gun signal for DM and providing the tantalizing opportunity to extract the properties of DM. Thereby one could use neutron star observations of unusual masses and radii to constrain the DM mass and to explore the stiffness of the DM equation of state.
 
\begin{acknowledgments}
We thank Laura Tolos, Mikel F. Barbat and Jan-Erik Christian for helpful discussions and comments. We would particularly like to thank the late Stephan Wystub for his help and advice. The authors acknowledge support by the Deutsche Forschungsgemeinschaft (DFG, German Research Foundation) through the CRC-TR 211 'Strong-interaction matter under extreme conditions'– project number 315477589 – TRR 211. 
\end{acknowledgments}

\section{Appendix}

Here we present additional results for the case of EoS2 to compare with the case of EoS1.

\begin{figure*}[]
    \centering
    \resizebox{1.0\textwidth}{!}{
    \begin{tabular}{cc}
        \subfloat[]{%
            \includegraphics[width=0.5\textwidth]{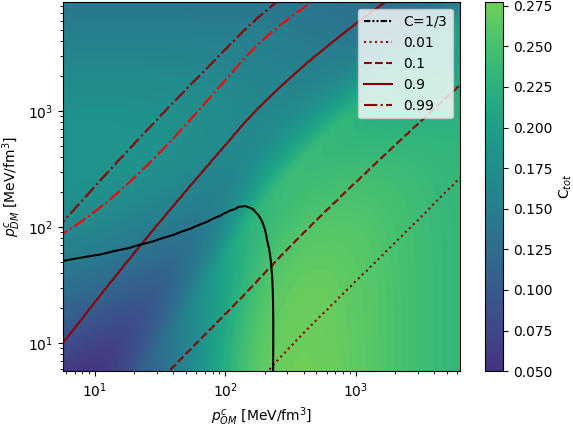}
            \label{fig:ContourEoS2a}
        } &
        \subfloat[]{%
            \includegraphics[width=0.5\textwidth]{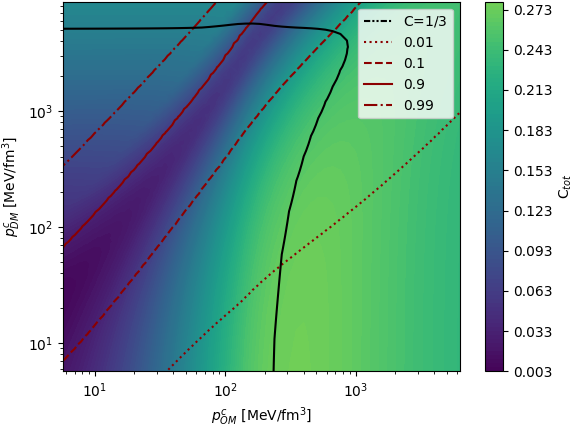}
            \label{fig:ContourEoS2b}
        } \\
        \subfloat[]{%
            \includegraphics[width=0.5\textwidth]{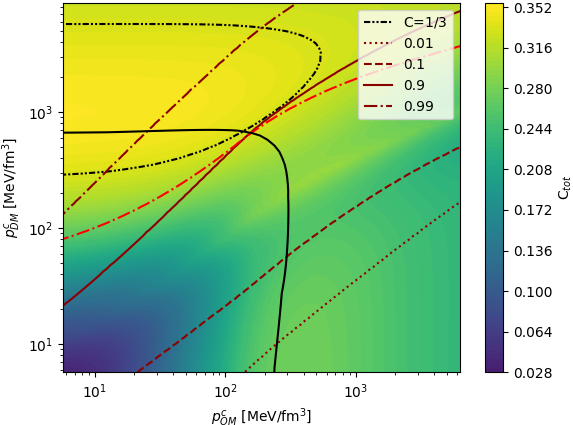}
            \label{fig:ContourEoS2c}
        } &
        \subfloat[]{%
            \includegraphics[width=0.5\textwidth]{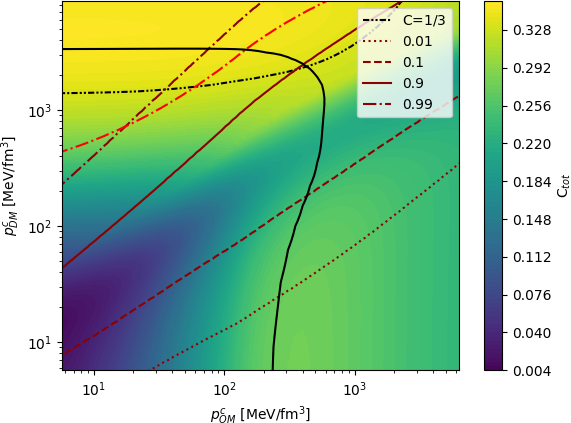}
            \label{fig:ContourEoS2d}
        }
    \end{tabular}}
    \caption{Contour-plots of EoS2 together with $n=4$ (upper panel) and $n=40$ lower panel, labeling as in figure \ref{fig:ContourEoS1}.}
    \label{fig:ContourEoS2}
\end{figure*}

In figure \ref{fig:ContourEoS2} we show contour-plots for the DM and OM central pressure with the compactness and the DM fractions. In figure \ref{fig:ContourEoS2a} the scenario for $n=4$ and a DM mass $m_b=300$ MeV is shown. Here the maximum compactness is $C=0.27$ which is determined by EoS2 since the maximum compactness of the $\phi^4$-potential is $C_{max}=0.16$ \cite{PauAmaro-Seoane_2010}. The stable configurations are below the black solid line, the unstable ones are above, for all plots. The region with stable objects compared to figure \ref{fig:ContourEoS2b} is relatively small and the case for a DM fraction of $99 \%$ is above the stability line. This changes for figure \ref{fig:ContourEoS2b} with a boson mass of $m_b=1000$ MeV and $n=4$. Here all DM fractions lie within the stable region. In figure \ref{fig:ContourEoS2c}, the case for $n=40$ and $m_b=200$ MeV, the compactness is now getting close to the limit of causality $C = 0.354$. In the case of a pure boson star the maximum compactness for this potential is $C=0.34$ which is reached in the case of high DM central pressures. The region of interest lies between the stability line and the line for $C=1/3$ since this marks the area where we have stable, ultra-compact neutrons stars. To generate these objects it is sufficient to have small OM central pressure of only a few MeV/fm$^3$. The same applies to figure \ref{fig:ContourEoS2d} where we have a boson mass of $m_b=300$ MeV. The difference between figure \ref{fig:ContourEoS2c} and figure \ref{fig:ContourEoS2d} is that for $m_b=300$ MeV the case for $90 \%$ DM is within the area of interest, whereas for $m_b=200$ MeV it lies slightly outside. These neutron star configurations are light with a small radius and a high compactness. However, the energy density of the OM is small so we do not expect a phase transition to quark matter in the core.


%

\end{document}